%% file: paper.tex
\documentclass[usenatbib]{mn2e}
\input{psfig}
\input{macros}
\begin{document}

\title[Galaxy Weak Lensing up to $z \sim 0.6$]{
Describing galaxy weak lensing measurements
from tenths to tens of Mpc
and  up to $z \sim 0.6$
with a single model
}
\author[Cacciato et al.]
       {\parbox[t]{\textwidth}{
        Marcello Cacciato$^{1}$\thanks{
        E-mail: cacciato@strw.leidenuniv.nl}, 
        Edo van Uitert$^{2}$, 
        Henk Hoekstra$^{1}$
        } \\
           \vspace*{3pt} \\
         $^1$Leiden Observatory, Leiden University, Niels Bohrweg 2, NL-2333 CA Leiden, The Netherlands\\
         $^2$Argelander-Institut f\"ur Astronomie, Auf dem H\"ugel 71, 53121 Bonn, Germany
           }


\date{}

\pagerange{\pageref{firstpage}--\pageref{lastpage}}
\pubyear{2012}

\maketitle

\label{firstpage}


\begin{abstract}
The clustering of galaxies and the matter distribution around them can
be described using the halo model complemented with a realistic description  of the way 
galaxies populate dark matter haloes. This has been used successfully to
describe statistical properties of samples of galaxies at $z<0.2$. Without
adjusting any model parameters, we compare the predicted weak lensing
signal induced by Luminous Red Galaxies to measurements from SDSS DR7
on much larger scales (up to $\sim 90 \, h_{70}^{-1}$Mpc)
and at higher redshift ($z \sim 0.4$). We find excellent agreement,
suggesting that the model captures the main 
properties 
of the galaxy-dark matter connection.
To extend the comparison to lenses at even higher redshifts we
complement the SDSS data with shape measurements from the deeper RCS2,
resulting in 
precise lensing measurements 
for lenses up to $z\sim 0.6$. 
These measurements are also well described using
the same model. 
Considering solely these weak lensing measurements,
we robustly assess that, up to $z \sim 0.6$, the number of
central galaxies as a function of halo mass is well described by a
log-normal distribution with scatter $\sigma_{\log L_{\rm c}}=0.146\pm0.011$, 
in agreement with previous independent studies at lower redshift.
Our results demonstrate the value of complementing the information about 
the properties of the (lens) galaxies provided by SDSS with deeper, high-quality imaging data.  
\end{abstract}


\begin{keywords}
galaxies: halos ---
large-scale structure of Universe --- 
dark matter ---
gravitational lensing ---
methods: statistical
\end{keywords}


\section{Introduction}
\label{sec:intro}

Since the advent of large and homogeneous galaxy surveys, it has
become possible to constrain the relation between the observed properties of
galaxies and their host dark matter haloes with ever increasing precision, 
albeit in a statistical sense. In particular, studies of
the observed abundances and clustering properties of galaxies
\citep[e.g.][]{ValeOstriker2004, Conroy2006, Shankar2006,
  ValeOstriker2006, Yang2008a, Moster2010, GuoWhiteBK2010,
  BehrConWech2010, Moster2012, Guzzo2000, Norberg2001, Norberg2002a,
  Zehavi2005, Wang2007} have played a crucial role in establishing
this relation with increasing detail.

Complementing these methods, the weak gravitational lensing signal
around galaxies of different observed properties (galaxy-galaxy
lensing) has emerged as another powerful technique to constrain this
relation. Since the first detections
\citep[e.g.][]{Brainerd1996,Griffiths1996,Hudson1998}, the
galaxy-galaxy lensing signal is now detected routinely as a function
of the properties of the lens galaxies, thanks to multi-wavelength
data becoming readily available \citep[e.g.][]{Fischer2000,
  McKaySheldonEtAl2001, Hoekstra2003, Hoekstra2005, Sheldon2004,
  Mandelbaum2006, Heymans2006, Parker2007, Mandelbaum2008,
  vanUitert2011,Choi12}. The systematics involved in these
measurements have also been studied in great detail \citep[see
  e.g.][]{Mandelbaum05, Mandelbaum2006, Mandelbaum2008}.  The main
application remains the study of the galaxy-dark matter connection,
such as measurements of scaling relations between halo mass and
baryonic properties \citep[e.g.][]{Hoekstra2005, Mandelbaum2006, Cacciato2009, Alexie10, vanUitert2011, Choi12}, constraints on the halo properties
\citep[e.g.][]{ Hoekstra2004, Mandelbaum2006, Limousin2007,
  Mandelbaum2008, vanUitert2012}, and measurements of bias parameters
\citep[e.g.][]{Hoekstra01, Hoekstra02, Sheldon2004}. More recently 
galaxy-galaxy lensing
has also been used as a cosmological probe in combination with galaxy
abundance and/or clustering measurements \citep[][]{CosmoCLFII,
  CosmoCLFIII, Rozo10, ZuWeinberg12, MandelCosmo12}.

There is growing scientific interest in probing the cosmic evolution
of structure formation in the Universe, which is now becoming possible
thanks to new and forthcoming galaxy surveys. For instance, one can
perform statistically representative analyses up to $z\sim 1$ in the
near future (e.g. KiDS \citealt{KiDS}, VIPERS
\citealt{VipersMarchetti}, Pan-STARRS \citealt{PSK},
DES\footnote{https://www.darkenergysurvey.org}, HSC\footnote{
  http://www.naoj.org/Projects/HSC/HSCProject.html}), and possibly up
to $z \sim 2$ in a decade, e.g. with missions such as
LSST\footnote{http://www.lsst.org}, and
Euclid\footnote{http://www.euclid-ec.org} \citep[][]{Euclid2011}.

Alongside the progress in observational capabilities, theoretical
modelling has also improved substantially. Numerical simulations have
proven important to investigate the link between galaxy-galaxy lensing
and the galaxy-dark matter connection\citep[e.g.][]{Tasitsiomi2004,
  LimousinKneibNatarajan2005, NatarajanDeLuciaSpringel2007,
  HayashiWhite2008}. Furthermore, the observed abundance, clustering
and lensing signal have been successfully explained using a
statistical description of the dark matter distribution in the
Universe as provided by the halo model  \citep[see e.g.][]{CoorayShethHaloModelRev,CosmoCLFI} coupled to a realistic model that describes
the way galaxies of different observable properties populate host
haloes \citep[see e.g.][]{Yang2003,CoorayMilo05,Cooray2006,Yang2008a}.

In this  study, we examine  the modeling of the  galaxy-galaxy lensing
signal up to $z \sim 0.6$. To this  end, we first compare a model that
describes the statistical properties  of galaxies at low redshift (van
den  Bosch et  al. 2013,  Cacciato et  al. 2013)  to  existing
galaxy-galaxy lensing  data measured  around Luminous Red  Galaxies at
higher redshift \citep{MandelCosmo12}. To extend the redshift range
even further, and to obtain higher precision measurements, we follow
\citet{vanUitert2011} and complement the
ninth data  release (hereafter  DR9) of the  Sloan Digital  Sky Survey
(hereafter SDSS) with $\sim 450$ square degrees of high-quality imaging data from the
second generation Red-sequence Cluster Survey (RCS2, \citealt{RCS2}).

This paper is organized as follows. We describe the analytical model
in \S2, its application to Luminous Red Galaxies in \S3.  We then
describe the surveys and the strategy to extract the new lensing
measurements in \S\ref{sec:data}.  Results are presented in
\S\ref{sec:results}.  Conclusions are drawn and discussed in
\S\ref{sec:conclusions}.

Throughout this paper, we adopt the most basic (`vanilla')
$\Lambda$CDM cosmological model.  Such $\Lambda$CDM cosmologies are
described by 5 parameters: the energy densities (in terms of the
critical density) of baryons, $\Omega_\rmb$, and cold dark matter,
$\Omega_{\rm dm}$; the spectral index, $n$, and normalization,
$\sigma_8$, of the initial power spectrum; and the Hubble parameter,
$h_{70} \equiv H_0/(70 \kmsmpc)$.  The flat geometry implies that
$\Omega_\Lambda = 1 - \Omega_\rmm = 1 - \Omega_\rmb - \Omega_{\rm
  dm}$. Throughout the paper, following the results of Cacciato et
al. (2013), we assume $(\Omega_\rmm,\Omega_\Lambda, \sigma_8,h_{70},
n, \Omega_\rmb h^2) = (0.278,0.722, 0.763,1.056,0.978, 0.0228)$.
Radii and densities are in comoving units\footnote{We write the mean
  density of the Universe as $\bar{\rho}_{\rmm} = \Omega_{\rm
    m}{\rho_{\rm crit}}$.}.  When physical units are used they are
explicitly indicated with `${\it p-}$'.  Furthermore, log is used to
refer to the 10-based logarithm.

\section{Modelling Galaxy-Galaxy Lensing}
\label{sec:model}

In this section we briefly describe how model predictions for the
galaxy-galaxy (hereafter g-g) lensing signal can be provided once one
has a statistical description of dark matter properties (i.e. their
average density profile, their abundance, and their large scale bias)
complemented with a statistical description of the way galaxies of a
given luminosity populate dark matter haloes of different masses 
(also known as halo occupation statistics).  The model is identical to the one
presented in \cite{CosmoCLFI} and successfully applied to SDSS in
\citet[hereafter C13]{CosmoCLFIII}. Readers familiar with this model
may skip this section and continue from \S\ref{sec:gglensingLRGS}
where we describe its application to Red Luminous Galaxies.

Weak gravitational lensing is sensitive to the mass distribution 
projected along the line-of-sight. Specifically, the quantity of interest is the
excess surface density (ESD) profile, $\Delta \Sigma(R)$, given by
\begin{equation}\label{shear} 
\Delta\Sigma(R,{\bar z}_{\rm le})= {2\over R^2} \int_0^R
\Sigma(R',{\bar z}_{\rm le}) \, R' \, \rmd R' - \Sigma(R,{\bar z}_{\rm
  le}).
\end{equation}
Here $\Sigma(R,{\bar z}_{\rm le})$ is the projected surface mass
density, which is related to the galaxy-dark matter cross correlation,
$\xi_{\rm gm}(r,{\bar z}_{\rm le})$, according to
\begin{equation}\label{Sigma_approx}
\Sigma(R,{\bar z}_{\rm le}) = \bar{\rho}_\rmm \int_{0}^{\omega_{\rm so}} 
\left[1+\xi_{\rm gm}(r,{\bar z}_{\rm le})\right] \, 
\rmd \omega
\,, 
\end{equation}
where the integral is along the line of sight with $\omega$ the 
comoving distance from the observer. The three-dimensional 
comoving distance $r$ is related to $\omega$ through $r^2 = \omega_{\rm le}^2 
+ \omega^2 - 2 \omega_{\rm le}\omega \cos \theta$. Here, $\omega_{\rm le}$ is the 
comoving distance to the lens, and $\theta$ is the angular separation between lens and 
source (see Fig.1 in \citealt{Cacciato2009}).
Note that the galaxy-dark matter cross correlation is evaluated at the
{\it average} redshift of the lens galaxies, ${\bar z}_{\rm le}$.

Observationally the ESD profile is inferred by measuring the average
tangential distortion of background galaxies (sources) around
foreground galaxies (lenses):

\begin{equation}\label{eq:gammat}
\langle\gamma_{\rm t}\rangle(R) = 
\frac{\Delta\Sigma(R)}{\Sigma_{\rm crit}}, \label{eq:dsigma}
\end{equation}

\noindent where $ \langle ... \rangle$ indicates the azimuthal average
inside an annulus at distance $R$ from the centre of the lens and of
width ${\rm d}R$. 
In Eq.(\ref{eq:gammat}),
$\Sigma_{\rm crit}$
is a geometrical factor determined by the distances of (lens and
source) galaxies:
\begin{equation}
  \Sigma_{\mathrm{crit}}=\frac{c^2}{4\pi G}\frac{D_{\rm so}}{D_{\rm
      le}D_{\rm le-so}(1+z_{\rm le})^2},
\end{equation}
with $D_{\rm le}$, $D_{\rm so}$, and $D_{\rm le-so}$ the angular
diameter distance to the lens, the source, and between the lens and
the source, respectively, and the factor $(1+z_{\rm le})^{-2}$
accounts for our use of comoving units.

Under the assumption that each galaxy resides
in a dark matter halo, $\Delta\Sigma(R,z)$ can be computed using a
statistical description of how galaxies are distributed over dark
matter haloes of different mass \citep[see e.g.][]{CosmoCLFI}.
Specifically, it is fairly straightforward to obtain the two-point
correlation function, $\xi_{\rm gm}(r,z)$, by Fourier transforming the
galaxy-dark matter power-spectrum, $P_{\rm gm}(k,z) $, i.e.
\begin{eqnarray}\label{xiFTfromPK}
\xi_{\rm gm}(r,z) = {1 \over 2 \pi^2} \int_0^{\infty} P_{\rm gm}(k,z) 
{\sin kr \over kr} \, k^2 \, \rmd k\,,
\end{eqnarray}
with $k$ the wavenumber.
$P_{\rm gm}(k,z)$, can be expressed
as a sum of a term that describes the small scales (one-halo, 1h),
 and one that describes the large scales
(two-halo, 2h),
 each of which can be further subdivided based
upon the type of galaxies (central or satellite) that contribute to
the power spectrum, i.e.,
\begin{equation}
P_{\rm gm}(k) = P^{\rm 1h}_{\rm cm}(k) + P^{\rm 1h}_{\rm sm}(k) 
+ P^{\rm 2h}_{\rm cm}(k) + P^{\rm 2h}_{\rm sm}(k)\,.
\end{equation} 
As shown in \citet{CosmoCLFI}, these terms can be written in
compact form as
\begin{equation}\label{P1h}
P^{\rm 1h}_{\rm xy}(k,z) = \int \calH_\rmx(k,M,z) \, \calH_\rmy(k,M,z) \, 
n_{\rm h}(M,z) \, \rmd M,
\end{equation}
\begin{eqnarray}\label{P2h}
\lefteqn{P^{\rm 2h}_{\rmx\rmy}(k,z) =
\int \rmd M_1 \, \calH_\rmx(k,M_1,z) \, n_{\rm h}(M_1,z) } \nonumber \\
& & \int \rmd M_2 \, \calH_\rmy(k,M_2,z) \, n_{\rm h}(M_2,z) \,
Q(k|M_1,M_2,z)\,,
\end{eqnarray}
where `x' and `y' are either `c' (for central), `s' (for satellite),
or `m' (for matter), $Q(k|M_1,M_2,z)$ describes the power
spectrum of haloes of mass $M_1$ and $M_2$, 
and it contains the large scale bias of haloes as well as a treatment of halo exclusion. 
Furthermore,
$n_{\rm h}(M,z)$ is the halo mass function of \citealt{Tinker2010}
\citep[see][for further detail]{CosmoCLFI, CosmoCLFIII}.
Here, we have defined
\begin{equation}\label{calHm}
\calH_\rmm(k,M,z) = {M \over \bar{\rho}_{\rmm}} \, \tilde{u}_\rmh(k|M,z)\,,
\end{equation}
\begin{equation}\label{calHc}
\calH_\rmc(k,M,z) = \calH_\rmc(M,z) = 
{\langle N_\rmc|M \rangle \over \bar{n}_{\rmg}(z)} \,,
\end{equation}
and
\begin{equation}\label{calHs}
\calH_\rms(k,M,z) = {\langle N_\rms|M \rangle \over \bar{n}_{\rmg}(z)} \,  
\tilde{u}_\rms(k|M,z)\,.
\end{equation}
Here $\langle N_\rmc|M \rangle$ and $\langle N_\rms|M \rangle$ are the
average number of central and satellite galaxies in a halo of mass $M
\equiv 4 \pi(200 {\bar \rho})R_{200}^3/3$, whereas $
\bar{n}_{\rmg}(z)$ is the number density of galaxies at redshift $z$.
We compute these quantities using the following expressions:
\begin{eqnarray}\label{eq:hod}
\langle N_{\rm x}|M \rangle &=&  
\int_{L_-}^{L_+} \Phi_{\rm x}(L|M) \, {\rm d} L ,
\end{eqnarray}
where $\Phi_{\rm x}(L|M)$ is the conditional luminosity function (see below and Appendix~A), $L_-$ and $L_+$ refer to the lower and upper limit of a luminosity bin, 
respectively. Again, the subscript `x' stands for either `c' (centrals) or `s'
(satellites), and
\begin{eqnarray}
\bar{n}_{\rmg}(z) &=& 
\int \langle N_{\rm g}|M \rangle n_{\rm h}(M,z) {\rm d} M \, .
\end{eqnarray}

Furthermore, $\tilde{u}_\rms(k|M)$ is the Fourier transform of the
normalized number density distribution of satellite galaxies that
reside in a halo of mass $M$, and $\tilde{u}_\rmh(k|M)$ is the Fourier
transform of the normalized density distribution of dark matter within
a halo of mass $M$. In this paper, supported by the results of
\cite{CosmoCLFIII}, we assume for both these profiles the functional
form suggested in Navarro, Frenk \& White (1997) . The conditional luminosity function ($\Phi_{\rm x}(L|M)$, hereafter CLF)
describes the \emph{average} number of galaxies with luminosities in
the range $L \pm \rmd L/2$ that reside in a halo of mass $M$.
Following \cite{CosmoCLFIII}, we parametrize the CLF with nine
parameters (see Appendix~\ref{sec:clf} for a thorough description). We
note here that the CLF methodology describes the halo occupation
statistics of \emph{both} central and satellite galaxies and it is not
limited to the choice of specific luminosity bins, rather it applies
to galaxies as a function of their luminosity.  This will be of
crucial importance when we will interpret the data presented in
\S\ref{sec:data}.

\subsection{Additional lensing terms}

In the analytical model used by C13, which was summarized above, the 
lensing signal is modelled as the sum of four terms: two describing
the small (sub-Mpc) scale signal mostly due to the dark matter density
profile of haloes hosting central and satellite galaxies; and the
other two describing the large (several Mpc) scale signal due to the
clustering of dark matter haloes around central and satellite
galaxies, respectively. This reads:
\begin{eqnarray}\label{eq:esdterms}
\Delta \Sigma(R,z) &=& \Delta \Sigma^{\rm 1h}_{\rm cm}(R,z)+\Delta \Sigma^{\rm 1h}_{\rm sm}(R,z)
\nonumber \\
& + & \Delta \Sigma^{\rm 2h}_{\rm cm}(R,z)+\Delta \Sigma^{\rm 2h}_{\rm sm}(R,z) \, .
\end{eqnarray}
In the halo model
the small scale signal (the 1-halo term) has two more contributors
corresponding to: i) the baryonic mass of the galaxies themselves; and
ii) the dark matter density profile of the sub-haloes which host
satellite galaxies.

The smallest scales probed by the data in this study are about 50 kpc,
which are much larger than the typical extent of the baryonic content
of a galaxy. Therefore, it is adequate to model the lensing signal due
to the baryonic content of the galaxy as the lensing due to a point
source of mass $M_{\rm g} \approx M_{\rm star}$ (see
e.g. \citealt{Alexie10}). This reads
\begin{equation}
\Delta \Sigma^{\rm 1h, g}(R,z) \approx \frac{\langle M_{\rm star}(z_{\rm le}) \rangle_{L_-}^{L_+} }{\pi \, R^2} \, .
\end{equation}
When accounting for the baryonic mass, this term adds to the other
four indicated in eq.(\ref{eq:esdterms}). Throughout the paper, we
model the lensing signal as in \S\ref{sec:model}. However, when
describing Figure~\ref{fig:modeldatacomparison}, we comment on how
model predictions are modified once the baryonic mass is taken into
account in the simplified way described above.  To that aim, we use
the value of the average stellar masses, $\langle M_{\rm
  star}\rangle$, for the galaxies in the luminosity bins under
investigation here (see \S5).  For completeness, we list these values
in Table~1.

The modeling of the dark matter density profile of the sub-haloes
which host satellite galaxies is conceptually simple (see e.g. 
\citealt[][]{Mandelbaum05,Li09, Giocoli10, Li12, RodriguezPuebla2013}) 
However, a
proper implementation of this term is hampered by the poor knowledge
of the subhalo mass function (see e.g. Giocoli, Tormen \& van den
Bosch 2008) and of the stripping mechanism \citep[see e.g.][]{Gao2004}
which occurs once a dark matter halo enters a larger halo, i.e. when
an initially central galaxy becomes a satellite. Many of the results
about such subhalo properties are obtained from pure N-body
simulations for which the limited mass resolution may still be a
important limiting factor. Furthermore, it is unclear how these
results are affected by various baryonic processes in place during
galaxy evolution \citep[e.g.][]{vanDaalen2011}.  Given these uncertainties
and since subhaloes only contribute a small fraction to the total
lensing signal on small scales (see e.g. \citealt{Li09}), 
in this paper, we refrain from modelling the lensing
term due to the subhaloes which host satellite galaxies. We comment on
the impact of this simplification when comparing model predictions
with actual measurements.

\section{SDSS lensing signal around LRGs}
\label{sec:gglensingLRGS}
\label{sec:submodel}
\begin{figure}
\psfig{figure=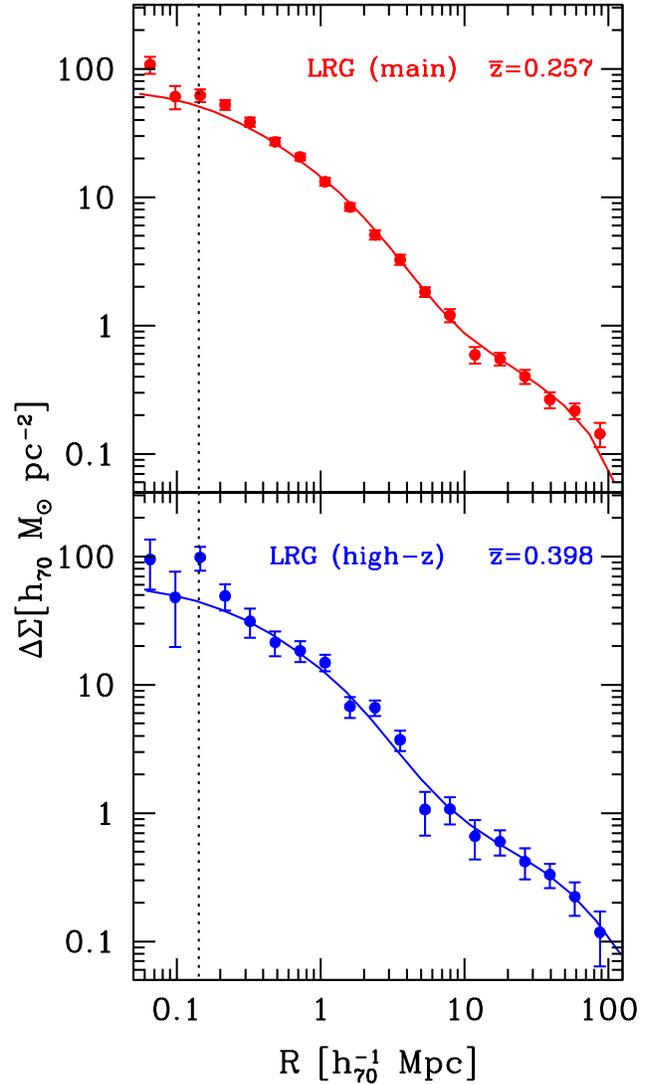,width=0.47\hdsize}
\caption{The excess surface density of LRGs from the analysis of SDSS
  DR7 by \citet{MandelCosmo12}. Solid lines refer to the model
  predictions for the lensing signal using the model parameters
  retrieved in C13, \emph{without further adjustment}. Note that the
  lensing measurements are uncertain at scales smaller than about 0.14
  $h_{70}^{-1}$ Mpc as indicated by the dotted vertical line.  }
\label{fig:modeldataLRG}
\end{figure}
The model summarized in \S2
was used by C13 to fit
the galaxy-galaxy lensing signal measurements performed via SDSS in
the spatial range $0.05 \lsim R \lsim 2 $ Mpc and at redshift $z\lsim
0.2$. The same model can be used to make predictions about the scale and redshift
dependence of the lensing signal. To test the robustness of the model,
it is therefore interesting to examine how it performs, without any
adjustments of the parameters (including the best-fit cosmology from
C13), when compared to different data.

We first consider the g-g lensing signal for a sample of Luminous Red
Galaxies \citep[LRGs,][]{Eisenstein01}. \citet{MandelCosmo12} have
measured the lensing signal around two LRG samples based on the SDSS
DR7 catalogue. The selection of LRGs allows the study of the dark
matter distribution via weak gravitational lensing at higher redshift
compared to the main sample. 
The effective redshifts of the two samples are
$z_{\rm le}
\approx 0.26$ and $z_{\rm le} \approx 0.40$. Both samples have
absolute magnitude limits $-23.2 < M_g <-21.1$.  Note that
$k$-corrections and evolution corrections to convert $r$-band
magnitude to $M_g$ are taken from \citet{Eisenstein01}. More
details about the procedure to select LRGs can be found in 
\citet{Kazin2010} and in \citet{MandelCosmo12}.

\begin{figure*}
\psfig{figure=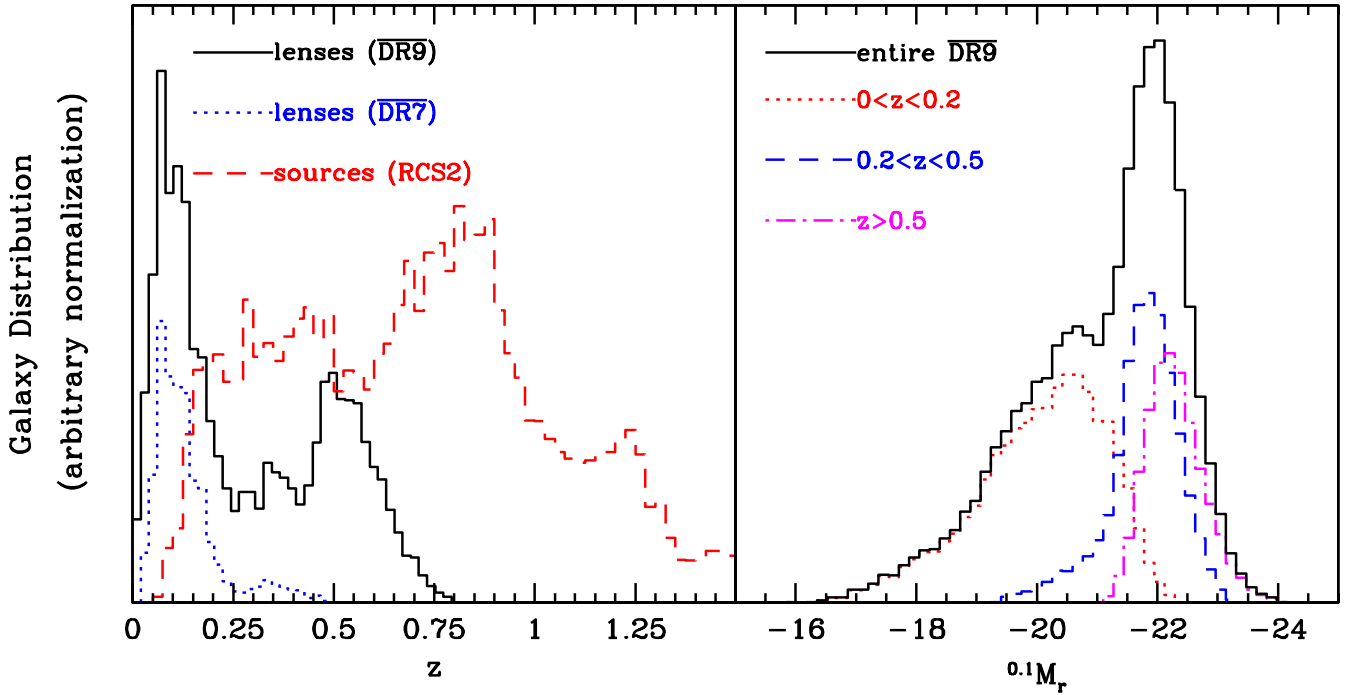,width=\hsize}
\caption{ {\it Left:} Redshift distributions of lens and source
  galaxies (with arbitrary normalization).  The black solid histogram
  show the lens distribution of the $\overline{DR9}$ sample used in
  this paper, whereas the blue dotted histogram refers to the lens
  distribution of $\overline{DR7}$ used by van Uitert et
  al. (2011). The red dashed histogram indicates the approximate redshift
  distribution of the source galaxies. {\it Right:} Distribution of
  absolute magnitudes of the lens galaxies. The black solid histogram
  refers to the entire $\overline{DR9}$ sample, whereas different
  lines refer to different subsamples defined via redshift cuts (see
  legend).}
\label{fig:zMrhist}
\end{figure*}

Figure~\ref{fig:modeldataLRG} shows the data (filled circles with
error bars). The model predictions (solid lines) are obtained by using
the same cuts as \citet{MandelCosmo12} and the \emph{same} model
parameters found in C13. The model, although constrained using the
main sample, describes the observed LRG signals very well. The value
of the reduced $\chi^2$ computed in the range\footnote{ Note that the
  LRG lensing measurements used here are uncertain at scales smaller
  than about 0.14 $h^{-1}_{70}$ Mpc (R. Mandelbaum private
  communication)} $0.2 \lsim R \sim 90 h^{-1}_{70}$ Mpc is 1.0 for the
main LRGs and 0.8 for the high-z LRGs. The lensing signal around LRGs
is reproduced over a large range of scales ($0.2 \lsim R \lsim 90
h_{70}^{-1}$ Mpc) and at high redshifts $z \sim 0.26$ and 0.4. 

It is worth emphasizing that the lensing signal on small and large
scales carries different information. To first order, smaller scales
probe the mass distribution within haloes, whereas larger scales probe
the cosmological framework (mostly through a combination of the
parameters $\Omega_{\rm m}$ and $\sigma_8$). We recall here that
\citet{CosmoCLFIII} embedded their analysis in a fully Bayesian
framework in which they also constrained the cosmological parameters
which define a `vanilla' $\Lambda$CDM cosmology.  They found that the
parameters $(\Omega_\rmm,\Omega_\Lambda, \sigma_8,h^{-1}_{70}, n,
\Omega_\rmb h^2) = (0.278,0.722, 0.763,1.056,0.978, 0.0228)$ best fit
their model. Hence the agreement with the measurements is an important
validation of the model determined by C13. It not only implies that
the parameters that describe the halo occupation distribution are also
valid at higher redshifts, but also that the cosmological parameters
are consistent.

Before comparing the model to g-g lensing measurements based on a
different data set in \S4, we exploit the quality of the agreement
between LRGs lensing data and model predictions to compute the average
host halo mass of LRGs for both the main and the high-z sample. The
estimation of the average halo mass follows from
\begin{eqnarray}\label{eq:averagehalomass}
\langle M_{200} \rangle & \equiv & \frac{1}{{\bar n_{\rm c}}(z_{\rm le})}
\int \langle N_{\rm c}|M \rangle n_{\rm h}(M,z_{\rm le})  M {\rm d} M \, , 
\end{eqnarray}
where $ n_{\rm h}(M,z_{\rm le})$ is the halo mass function \citep{Tinker2010}
at the lens redshift, and $\langle N_{\rm c}|M \rangle$ is computed via 
eq.~(\ref{eq:hod}). 
We find that both the main and the high-z LRGs reside in haloes with 
$\langle \log[M_{200}/(h^{-1}_{70} M_{\odot})] \rangle \approx 13.6$,
in general agreement with independent previous studies 
\citep[see e.g.][]{Mandelbaum2006, Zheng09LRGs}.

\section{RCS2 galaxy-galaxy lensing signal}
\label{sec:data}

\begin{figure}
\psfig{figure=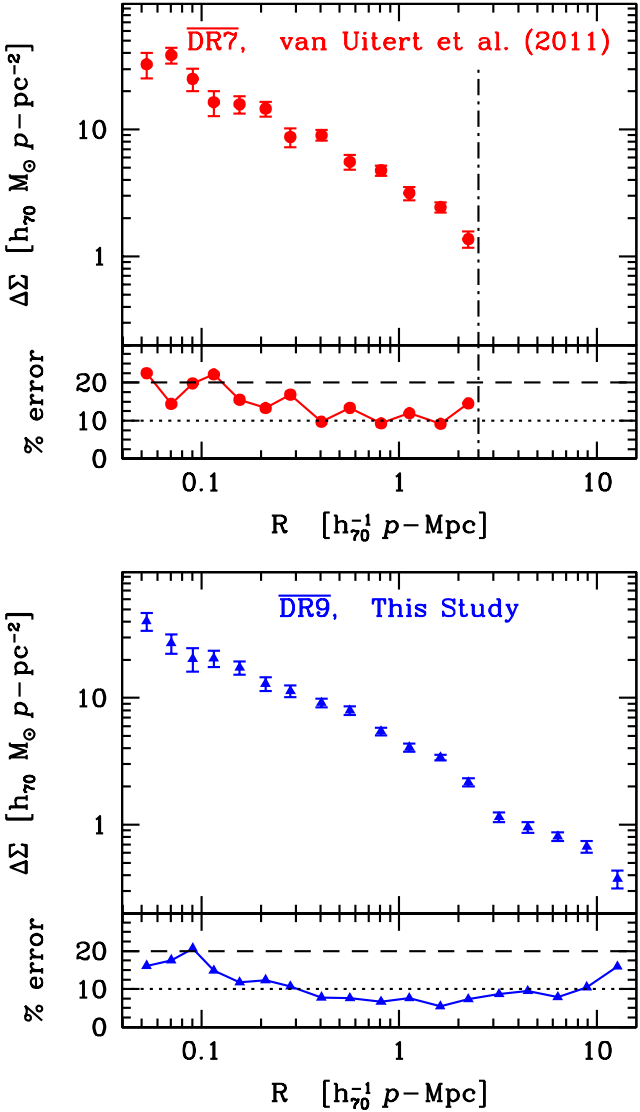,width=\hsize}
\caption{ {\it Upper panel.}  The ESD signal measured in Van Uitert et
  al. (2011). Note the limited spatial scale ($0.05 \lsim R \lsim 2$
  {\it p-}Mpc) and that the error budget is, {\it on average}, between
  10 and 20 \%.  {\it Lower panel.}  The ESD measurements presented in
  this study.  Note that the probed spatial range now extends up to $R
  \sim 10 $ {\it p-}Mpc and that the error budget is below 10 \% for
  most of the probed scales.  }
\label{fig:ESD-DR9}
\end{figure}
%

\begin{figure}
\psfig{figure=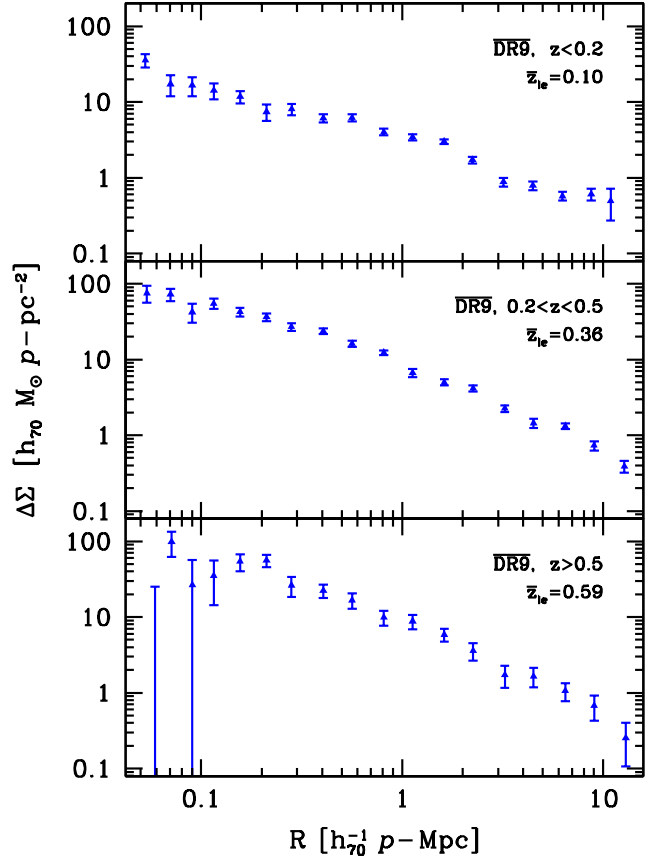,width=\hsize}
\caption{
The ESD measured in this study for three subsamples of $\overline{DR9}$: i) low $(z<0.2)$;
ii) intermediate $(0.2 < z < 0.5)$; and iii) high $(z >0.5)$ redshift. Note the high quality of the signal 
at both intermediate and high redshift, well beyond the regime probed by SDSS-alone studies.
}
\label{fig:ESD-DR9-2}
\end{figure}

The weak lensing signal decreases as the lens galaxy approaches the source.
This limits the g-g lensing analysis of the
SDSS main spectroscopic sample (DR7) to lenses with $z \sim 0.2$ when
SDSS shape measurements are used.  Reliable measurements at higher
redshift are only possible by targeting very luminous lens galaxies;
see the case of LRGs in the previous section. A unique aspect of the
SDSS is the wealth of spectroscopic information, which extends to
higher redshifts as is shown in Figure~\ref{fig:zMrhist}. To improve
the signal-to-noise ratio for higher redshift lenses, van Uitert et
al. (2011) 
measured the shapes of
source galaxies using imaging
data of a deeper survey that overlap with the SDSS. Specifically, \citet{vanUitert2011} studied the lenses in the
region of SDSS DR7 that overlaps with the second generation Red-sequence Cluster Survey (RCS2; \citealt{RCS2}). RCS2 is a 900
square degree imaging survey in three bands ($g'$, $r'$ and $i'$)
carried out with the Canada-France-Hawaii Telescope (CFHT) using the 1
square degree camera MegaCam. The RCS2 data are $\sim$2 magnitudes
deeper than the SDSS in the $r'$-band and the median seeing of 0.7$''$
is roughly half that of the SDSS. Consequently the survey is well
suited to improve the lensing constraints for lenses with $z\geq0.3$,
where the source distribution of the main sample in SDSS decreases
significantly. In particular, it allows us to improve the lensing
constraints for the most massive/luminous galaxies, which
are preferentially selected at higher redshifts.

As shown in Figure~\ref{fig:zMrhist}, a major change between the
spectroscopic samples of SDSS DR7 and DR9 is that high redshift
$(z>0.3)$ LRGs were targeted as part of the Baryon Oscillation
Spectroscopic Survey (BOSS; \citealt{Anderson2012}). As we show in this
section, the lensing signal around these lenses can be determined with
high precision using RCS2 shape measurements. 

The SDSS DR9 \citep{SDSSDR9} overlaps with 471 RCS2 pointings. This
amounts to roughly 450 square degrees (about 150 square degrees more
overlap than between the RCS2 and the DR7 used in Van Uitert et
al. 2011). The lens sample in the study presented here consists of all
objects from the DR9 in the overlapping area that have a reliable
spectroscopic redshift (according to the SciencePrimay flag) and that
are spectroscopically classified as galaxies.  In contrast to \citet{vanUitert2011}, we do not require the DR9 objects to have a
match with an object from the RCS2 catalogues, but only that they
reside within the field of view of a RCS2 pointing.  This leads to a
lens sample of $\sim 70,000$ objects, four times more than the lens
sample used in \citet{vanUitert2011}. In the remainder of the
paper, we shall refer to this sample as $\overline{DR9}$.

The redshift distribution of the lenses is shown in the left panel of
Figure~\ref{fig:zMrhist}.  For comparison, we also show the lens
sample used in \citet{vanUitert2011}, labelled as $\overline{DR7}$.
The majority of lenses with redshifts $z>0.3$ are LRGs, whereas at
lower redshifts, our lens sample consists of a mix of early-type and
late-type galaxies. We do not consider these samples separately in
this paper, as the halo model in use does not account for this split. The right-hand panel of Figure~\ref{fig:zMrhist} shows the
distribution of absolute magnitudes for the whole sample, and for
different redshift slices. The luminosities of the lenses are computed
using the $r$-band Petrosian magnitudes from the SDSS photometric
catalogues, corrected for extinction using the dust maps of \citet{Schlegel1998}. $K$ corrections were calculated to $z=0.1$ using the
{\tt KCORRECT} v4\_2 code \citep{Blanton2003a,BlantonRoweis07}. Finally, a passive luminosity evolution correction, $E$,  
was computed following \citet{MandelCosmo12}.
In summary, the absolute magnitudes were
computed as \mbox{$^{0.1}M_r=m+DM-K(z=0.1)+E$}, 
where $m$ is the apparent magnitude of a galaxy,
$DM$ is the distance modulus,
$K$ is the correction mentioned above, and
$E$ is the passive evolution correction taken from \citet{MandelCosmo12}, 
i.e. \mbox{$E=2(z-0.1)[1-(z-0.1)]$}.
We will comment on the impact of this assumption in \S5.2.1

The creation of the shape measurement catalogues for the RCS2 is
detailed in \citet{vanUitert2011} and we refer the reader to it for
a detailed description. Since we lack redshifts for the background
galaxies, we select galaxies with $22<m_{r'}<24$ that have a reliable
shape estimate\footnote{ We excluded sources with an ellipticity $>$1,
and those whose photometry was deemed unreliable (e.g. due to image
artefacts or neighbouring objects) as indicated by the flags of the
source extraction program {\tt SExtractor} (Betrin \& Arnouts
1996).}  as sources. The resulting average source density is
6.3/arcmin$^2$. The approximate source redshift distribution for the
sources (left-hand panel of Figure~\ref{fig:zMrhist}) is obtained by
applying identical magnitude cuts to the photometric redshift
catalogues of the COSMOS field \citep{IlbertCOSMOS}. This procedure is detailed in Appendix~C.

In contrast to \citet{vanUitert2011}, we do not limit our lensing
measurements to individual pointings, but we include the sources from
neighbouring patches when measuring the azimuthally averaged
tangential shear. This has the advantage that the lensing
signal-to-noise at large radii improves, due to the larger number of
sources at these separations. Another advantage of including
neighbouring patches is that it reduces the impact of systematic
contributions to the lensing signal, as is explained in Appendix~B.
There we also present a detailed description of the error estimate.

To compare the statistical power between the current work and the
analysis in \citet{vanUitert2011}, we show the lensing signal of
the respective lens samples in Figure~\ref{fig:ESD-DR9}.  We find that
the signal-to-noise of the lensing measurements improves by about 50
per cent on average. Importantly, the lensing signal is robustly
measured out to larger separations (see Appendix~B). We also split the
lens sample in three redshift bins, and show the lensing signals for
each bin in Figure~\ref{fig:ESD-DR9-2}. This illustrates that even at
$z>0.5$ (with ${\bar z}_{\rm le}\sim 0.59$), we are able to obtain
significant lensing measurements.  Furthermore, the higher
normalisation of the lensing signal measured at higher redshift is
indicative of the fact that, not surprisingly, more massive lenses
are selected at higher redshift.

\section{Results}
\label{sec:results}

The C13 model was constrained combining SDSS galaxy abundance and clustering 
measurements to g-g lensing data at low redshift, $z \lsim 0.2$, and relatively small scales, 
$R \lsim 2 $Mpc. The comparison with the LRG sample in \S3 provides an important test of
the model, but the lensing measurements are derived from the same
pipeline as the data used by C13. It is therefore interesting to
study how well the predictions compare to the results of the independent analysis 
that uses RCS2 shape measurements.
Such a comparison tests both the fidelity of the shape measurements \emph{and} 
the model at even higher redshifts. To do so, we split the lens
sample in eight luminosity bins and compare the g-g lensing signal
to model predictions (see \S5.1). Following this comparison we proceed to use the
lensing measurements to investigate the possibility to constrain 
the galaxy-dark matter connection at those higher redshift (see \S5.2).
 
\subsection{Comparison of the RCS2 lensing signal to model predictions}
\begin{table}\label{tab:esddata}
  \caption{Properties of the excess surface density data}
\begin{center}
\begin{tabular}{lcccc}
\hline\hline
 Label & ${\tilde M}_r$ & lg$M_\star$ 
 &${\bar z}_{\rm le}$ & $N_{\rm le}$ \\
  (1) & (2) & (3) & (4) & (5) \\
\hline
$L_1$ & $(-18.0,-17.0]$ & 9.21& 0.07 &  1,418 \\
$L_2$ & $(-19.0,-18.0]$ & 9.72& 0.09 &  3,650 \\
 $L_3$ & $(-20.0,-19.0]$ & 10.23& 0.12 &  8,918  \\
 $L_4$ & $(-21.0,-20.0]$ & 10.75& 0.19  & 15,254 \\
 $L_5$ & $(-21.5,-21.0]$ & 11.09& 0.36  & 14,013 \\
 $L_6$ & $(-22.0,-21.5]$ & 11.32& 0.44  & 13,555 \\
 $L_7$ & $(-22.5,-22.0]$ & 11.57& 0.51  & 5,730 \\
 $L_8$ & $(-23.0,-22.5]$ & 11.82& 0.59  & 1517 \\
\hline\hline
\end{tabular}
\end{center}
\medskip
\begin{minipage}{\hssize}
  The galaxy samples used to
  measure the excess surface density profiles, $\Delta\Sigma(R)$.
  For each of these samples column (1) lists the magnitude label, 
  column (2) lists the magnitude range, where 
  ${\tilde M}_r \equiv {^{0.1}M}_r$, 
   column~(3) lists the log of the average stellar mass
  (lg$M_\star \equiv \log{[\langle M_{\rm star}/h^{-1}_{70}M_{\odot} \rangle]} $), 
  column (4) lists the mean redshift, and
  column (5) lists the 
  number of lens galaxies.
\end{minipage}
\end{table}

\begin{figure*}
\psfig{figure=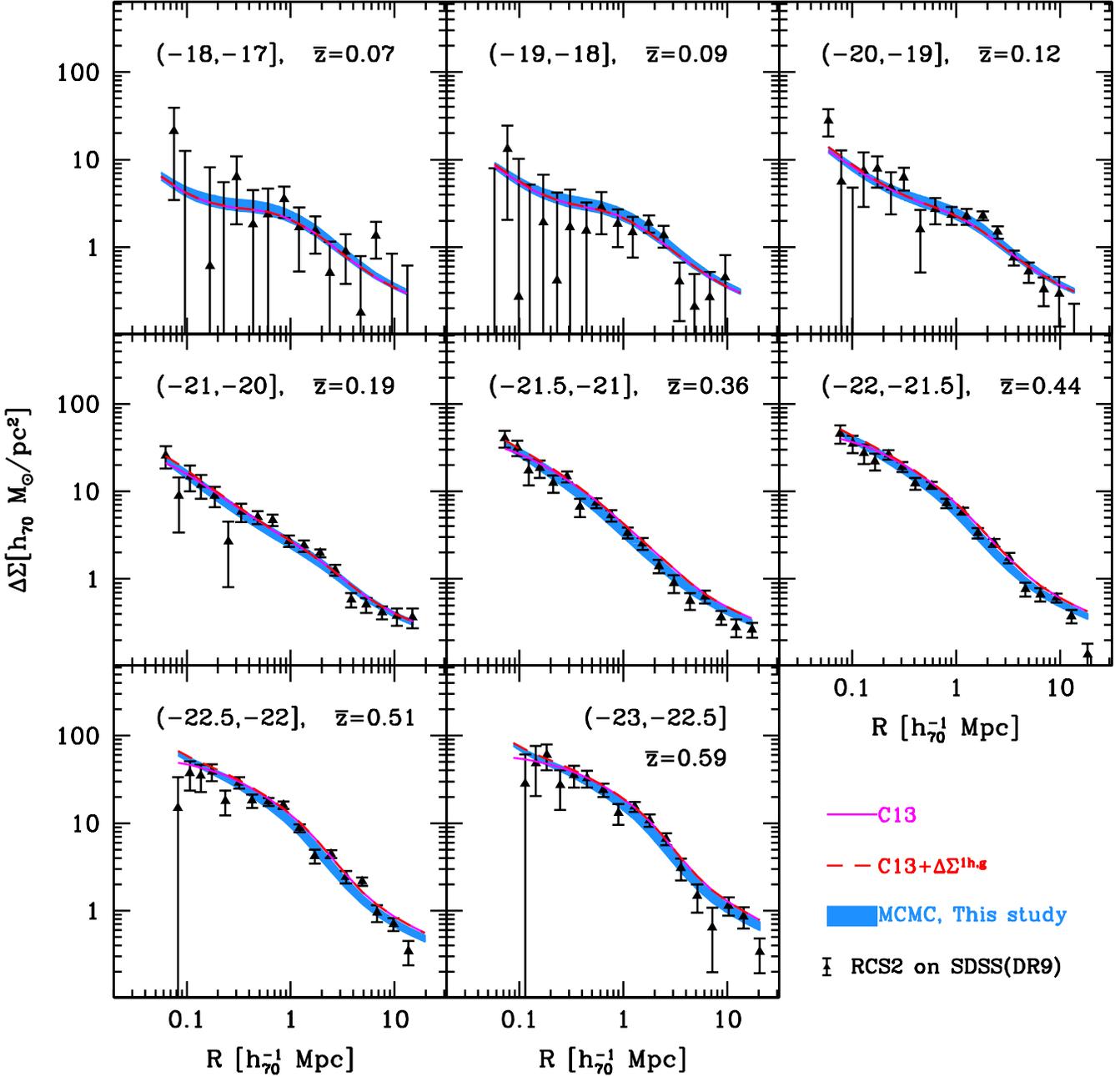,width=\hsize}
\caption{ The excess surface density around lenses in DR9 that overlap
  with RCS2. The black triangles indicate the lensing signal measured
  using the RCS2 imaging data. Magenta solid and red dashed lines
  refer to the model predictions from C13, see \S\ref{sec:model} and
  \S\ref{sec:submodel}, respectively.  The shaded cyan region refers
  to the MCMC results where we consider $L_0, M_1, \gamma_2$ and $\sigma_{\rm
    logLc}$ as free parameters.  }
\label{fig:modeldatacomparison}
\end{figure*}

To study the model predictions as a function of luminosity we divide
the $\overline{DR9}$ lens sample in eight luminosity bins. The main properties of
each bin are listed in Table~1. The black triangles with error bars in
Figure~\ref{fig:modeldatacomparison} indicate the resulting ESD
measurements based on the RCS2 lensing catalog as a function of
projected lens-source separation.  
Over the large range in luminosity and scale ($0.05\lsim
R\lsim10$ $h_{70}^{-1}$ Mpc) we measure a significant lensing signal.
Especially the measurements for the highest luminosity bins
represent an improvement over what can be done with SDSS
data alone. 

The magenta solid lines in Figure~\ref{fig:modeldatacomparison}
correspond to the model predictions based on the study by C13. Those
predictions are in overall agreement with 
the RCS2 measurements of the ESD on all scales,
including those well outside the range used to constrain the C13 model. 
The agreement between
model predictions and data supports the findings in C13 both in terms
of the halo occupation statistics \emph{and} the cosmological
parameters. Furthermore, the mutual agreement of the model with the
data obtained with SDSS alone \citep{Mandelbaum2006} and with those
obtained here using RCS2 implies overall consistency of the lensing signal
from the two surveys. Interestingly, for the highest four luminosity bins,
the C13 model predictions seem to systematically over-estimate the g-g lensing signal
at the $1 \sigma$ level. Although this aspect may be regarded as marginal given that 
the model predictions were not at all tuned to reproduce these data, we will 
further investigate the relevance of this slight disagreement in \S5.2.

Model predictions based on C13 can be easily modified to account for
the contribution to the lensing signal due to the galaxy baryonic mass
(red dashed lines in Figure~\ref{fig:modeldatacomparison}).  Using the
simplest assumption for how the stellar content of the galaxy may
contribute to the lensing signal (see \S\ref{sec:submodel}), we note
that fainter/less massive galaxies are virtually unaffected by such a
correction, whereas brighter/more massive galaxies exhibit a boost of
the lensing signal on scales $R \lsim 0.2 h_{70}^{-1}$ Mpc, reaching
up to a factor of about 1.5 on scales $R \sim 0.05 h_{70}^{-1}$ Mpc.

We note here that in the current analysis the average stellar mass per
luminosity bin is estimated by matching our lens sample to the MPA-JHU
DR7 value added catalogue\footnote{\tt
  http://www.mpa-garching.mpg.de/SDSS/DR7/} which provides stellar
mass estimates.  Specifically, we use the matching objects to fit a
linear relation between absolute magnitude and stellar mass, use this
relation to assign a stellar mass to all our lenses, and finally
determine the average for each luminosity bin (reported in Table~1).
A technical caveat must be mentioned here: the MPA-JHU catalogue only
contains galaxies from the DR7, and not the more recently observed
ones from BOSS. Therefore, when we use the average stellar mass to
compute the baryonic term in the halo model, we implicitly assume that
the relation between luminosity and stellar mass is similar for the
DR7 galaxies as for those that were observed as part of BOSS. This
assumption may not be accurate, but the use of the stellar mass in
this paper serves only to roughly quantify on which scales and by what
amount the g-g lensing signal might be affected by the baryonic mass of the
galaxy.

As a last cautionary note, we comment here on the fact that the model
presented in this paper does not account for the mass distribution in
the subhaloes which host satellite galaxies.  As the quality of the
lensing signal improves, it will become mandatory to add this extra
term especially if one aims to retrieve the amount of stellar mass by
fitting the small scale ($R \lsim 0.2 h_{70}^{-1}$Mpc) lensing signal.

\subsection{Constraining the galaxy-dark matter connection with weak lensing only}

More luminous galaxies reside {\it on average} in more
massive haloes. Using Eq.(\ref{eq:averagehalomass}) we find that this
is indeed the case, and that the luminosity bins listed in Table~1
correspond to halo masses that range from $\langle {\rm log}
M_{200}/(h_{70}^{-1} M_{\odot} \rangle \sim 11.4$, to $\langle {\rm
  log} M_{200}/(h_{70}^{-1} M_{\odot}) \rangle \sim 14.2$. The
signal-to-noise of the measurements presented in
Figure~\ref{fig:modeldatacomparison} is highest for lens galaxies with
$-23 \lsim {^{0.1}M}_r \lsim -21$, which correspond to relatively
massive haloes (12.5 \lsim $\langle {\rm log} M_{200}/(h_{70}^{-1}
M_{\odot}) \rangle \lsim 14.2$).
We explore here whether, thanks to the improved precision at the highest masses,
we can constrain the model parameters which govern this regime
using {\it solely} g-g lensing measurements. 
To this aim, we employ the same model used so far,
but we now leave the parameters that govern the massive
end of the galaxy luminosity-halo mass relation free to vary.  In the
parameterization used in this paper (see Appendix~A for more detail),
the relation between the luminosity of a central galaxy and its host
halo mass is assumed to be:
\begin{eqnarray}\label{eq:LcM}
L_\rmc(M) &=& L_0 {(M/M_1)^{\gamma_1} \over 
\left[1 + (M/M_1) \right]^{\gamma_1-\gamma_2}}
\nonumber \\
& \sim &
L_0 \left(\frac{M}{M_1}\right)^{\gamma_2} 
\, {\rm for}\, {M \gg M_1}
\, .
\end{eqnarray}
Furthermore, the {\it average} number of central galaxies of a given luminosity 
is related to the halo mass via a log-normal distribution:
\begin{eqnarray}
\Phi_\rmc(L|M) \,{\rmd}L &=& 
{\log\, e \over 
\sqrt{2\pi} \, \sigma_{\log L_{\rm c}}
} 
{\rm exp}
\left[
- { {(\log L -\log L_\rmc )^2 } 
\over
2\,\sigma_{\log L_{\rm c}}^2} \right]\, 
{\rmd L \over L}
\,,
\nonumber \\
\end{eqnarray}
where $\sigma_{\log L_{\rm c}}$ indicates the scatter in luminosity at
fixed halo mass and $\log L_\rmc$ is, by definition, the expectation
value for the logarithm of the luminosity of the central
galaxy:
\begin{equation}
\log L_\rmc = \int \Phi_\rmc(L|M) \, \log L \, {\rmd L} \, .
\end{equation}
Here, we consider $L_0, M_1$, $\gamma_2$, and $\sigma_{\rm logL_c}$ as four
free parameters, while keeping $\gamma_1$ fixed to 3.18, 
the value retrieved in C13. 
The first free parameter has the units of a luminosity ($h_{70}^{-2}
L_{\odot}$), the second has the units of a mass ($h_{70}^{-1}
M_{\odot}$), whereas the remaining two are dimensionless.

To determine the probability distribution of the model parameters
discussed above we run a Markov Chain Monte Carlo\footnote{The chain
  consists of four different chains which start from different initial
  guesses in the parameter space. In total, we perform about three
  million model evaluations. With an average acceptance rate of $\sim
  30\%$, the complete chain used in the analysis is a well converged
  chain of one million model evaluations.} 
(hereafter MCMC) using the standard Metropolis-Hasting algorithm 
\citep{Metropolis53}. In this chain, the parameters 
$L_0, M_1, \gamma_2$ and $\sigma_{\rm logL_c}$
are free to vary and no prior information is used, whereas the
remaining parameters are fixed at the same value as the one in the C13 model
(see also Appendix A).  
As the satellite fraction is supposed to be very low for bright galaxies 
\citep[e.g.][]{Mandelbaum2006,Cacciato2009,CosmoCLFIII,vanUitert2011}, 
and the faintest galaxies in this study have relatively large uncertainties, 
we do not expect significant biases from selecting a subsample of the model parameters 
that governs the galaxy-dark matter connection of central galaxies only.

\begin{figure}
\psfig{figure=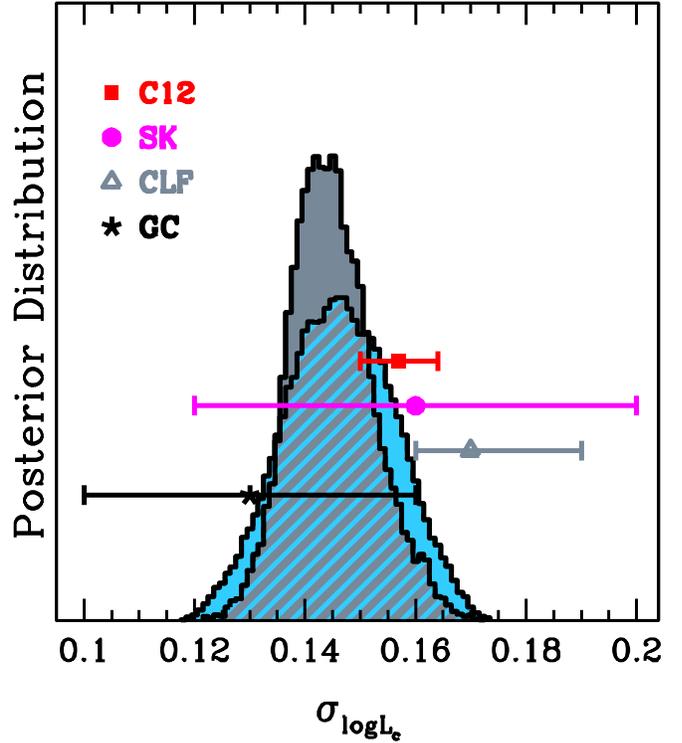,width=0.49\hdsize}
\caption{
Posterior distribution of the parameter $\sigma_{\log L_{\rm c}}$.
Blue shaded histogram refers to the analysis in \S5.1 and \S5.2,
whereas the grey shaded histogram refers to the result of the test performed in \S5.2.1
to assess the sensitivity of the analysis to passive evolution correction. 
The value of the scatter $\sigma_{\log L_{\rm c}}$ is robust against the uncertainties
deriving from passive evolution correction.
}
\label{fig:Scatter}
\end{figure}

The 95\% confidence levels of the g-g lensing models explored with the
MCMC are indicated by the cyan shaded regions in
Figure~\ref{fig:modeldatacomparison}. As expected, the subset of
parameters that we have varied has almost no impact on the predictions
for the lensing signal around the faintest galaxies. For brighter
galaxies, the MCMC brings the model in better agreement with the observables
than the initial C13 model predictions.
The agreement between model predictions and data has improved by assigning smaller 
halo masses to galaxies of the same luminosity. From our analytical model
(see especially eq. 17 and 18), one can see that lower halo masses at the same luminosity
can be obtained by altering the $L_{\rm c}(M)$ relation at 
the massive end or by increasing the scatter, $\sigma_{\log L_{\rm c}}$. 
As outcome of the MCMC we find that the $L_{\rm c}(M)$ relation has substantially 
changed from the one retrieved in C13. However, as discussed in the following subsection, 
the inference of the parameters which govern 
the $L_{\rm c}(M)$ relation is very sensitive to the assumed correction for luminosity evolution, which is uncertain.
Interestingly, the inference of the parameter $\sigma_{\log L_{\rm c}}$  is more robust 
against those uncertainties. Therefore, we report here only the corresponding result.
The blue shaded histogram in Fig.\ref{fig:Scatter} 
shows the posterior distribution of the scatter in the number of galaxies of a given
luminosity at any halo mass, $\sigma_{\rm logL_c}$ (see Appendix~A for
more details on this parameter). We
find that $\sigma_{\rm logL_c}$= $0.146\pm 0.011$ (median $\pm$ one
standard deviation), in excellent agreement with independent studies based
on abundance, clustering, and/or satellite kinematics at lower
redshift.  Specifically, using a large SDSS galaxy group catalogue,
Yang, Mo \& van den Bosch (2008) obtained $\sigma_{\rm
  logL_c}=0.13\pm0.03$ (black star) and they did not find evidence for
a halo mass dependence.  Cooray (2006) explicitly assumed no mass
dependence in $\sigma_{\rm logL_c}$ when studying the luminosity
function and clustering properties of SDSS galaxies, and found
$\sigma_{\rm logL_c}=0.17^{+0.02}_{-0.01}$ (grey triangle).
\citet{More2009a} studied the properties of satellite galaxy
kinematics around massive/luminous central galaxies and found
$\sigma_{\rm logL_c}=0.16\pm0.04$ (magenta circle). Finally, C13
combining abundance, clustering and lensing of galaxies in SDSS found
$\sigma_{\rm logL_c}=0.157\pm0.007$ (red square).

\subsubsection{Sensitivity to passive evolution correction}
\begin{figure}
\psfig{figure=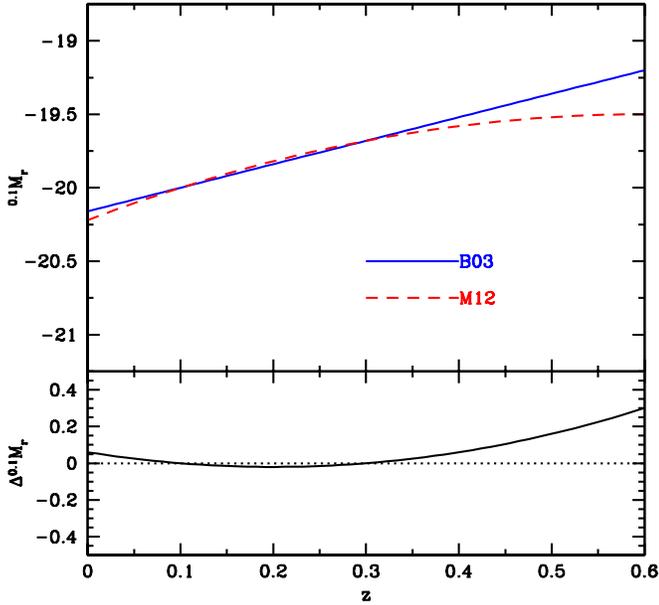,width=0.49\hdsize}
\caption{
{\it Upper panel.} The reference absolute magnitude, ${^{0.1}M_r} = {^{z}M_r} + E$,
as a function of redshift computed via the passive evolution correction, $E$, 
suggested by Blanton et al. (2003, blue solid line labelled B03) and Mandelbaum et al. 
(2012, dashed red line labelled M12) for the case of a galaxy with ${^{z}M_r}=-20$. 
Note that a galaxy with ${^{z}M_r}=-20$ at $z>0.1$ 
will be rescaled to a fainter reference magnitude, ${^{0.1}M_r}$, 
at the reference redshift $z=0.1$.
{\it Lower panel.} The difference between the B03 and M12 passive evolution corrections,
$\Delta {^{0.1}M_r} = {^{0.1}M_r^{B03}}-{^{0.1}M_r^{M12}}$.
}
\label{fig:PassiveEvolution}
\end{figure}
Ideally, one would like to compare the results on the $L_{\rm c}(M)$ relation
obtained here with those obtained at lower redshift to infer an evolutionary scenario. 
However, the physical interpretation of the
results of the MCMC is hampered by the fact that
the sample of galaxies used in this analysis is not uniform.
In fact, the mix of early- and late-type galaxies changes with redshift
due to the luminosity-based selection of the lenses. This leads to an 
uncertain
correction for luminosity evolution 
which enters in the definition of the reference absolute magnitude of a  
galaxy,
${^{0.1}M_r}$ (see \S4). 
In what follows, we will show how a small variation in the correction of 
luminosity evolution impacts the lensing analysis and the corresponding model 
parameters. 
This example highlights the sensitivity of our model to the intrinsic 
luminosity evolution of a galaxy.

In the analysis presented in \S5.1 and 5.2, we have used the passive 
evolution 
correction, $E$,  suggested in \citet{MandelCosmo12}. 
However, \citet{Blanton2003a} suggested a simpler functional form 
that has been widely adopted in the literature:
$E=1.6(z-0.1)$. The difference between these two 
functions is highest at higher redshift, 
reaching about 0.3 magnitude at the highest redshift 
of interest here, $z \sim 0.6$ (see Fig.~\ref{fig:PassiveEvolution}).
Using the \citet{Blanton2003a} expression for the passive evolution 
would have only a minor impact on the lensing analysis 
by \citet{MandelCosmo12} as LRGs are selected only up to $z\sim0.4$
and they are not split further into luminosity bins.
However, in our analysis lens galaxies are selected in narrow luminosity bins 
and up to $z\sim0.6$. A different selection of lens galaxies 
directly translates into a different lensing signal. 
Specifically, we find that using the passive evolution correction of \citet{Blanton2003a}
leaves the lensing signal of faintest galaxies virtually unaltered,
while leading to a higher normalisation (at about $1 \sigma$ level)
of the lensing signal of the four brightest bins.  
Repeating the MCMC analysis as in \S5.2 but on this new selection of lenses,
we recover similar values of the scatter,
$\sigma_{\log L_{\rm c}}$, as indicated by the grey shaded histogram in 
Fig.\ref{fig:Scatter}. However, we retrieve values for $M_1$, $L_0$, and $\gamma_2$
that differ by about 3 $\sigma$
from our initial analysis\footnote{We have checked 
that also this MCMC has converged.}. The changes in the retrieved model parameters are to be attributed to the expected degeneracies due to the 
assumed parametrization (see eq. [17]).
We conclude that the retrieved values of 
the model parameters $M_1$, $L_0$, and $\gamma_2$  are 
sensitive to the details of luminosity evolution correction. Therefore, 
in this paper we refrain from drawing any conclusion about the 
cosmic evolution of the $L_{\rm c}(M)$ relation,
deferring it to the analysis of a homogeneous sample of early-type galaxies 
(van Uitert et al., in preparation) for which passive luminosity correction 
can be modelled with higher confidence than in the current study
where we consider a mix of early- and late-type lens galaxies.

\section{Conclusions}
\label{sec:conclusions}

We investigated how measurements of the galaxy-galaxy lensing signal
around lenses at increasingly higher redshifts and at larger projected
distances can be used to study the galaxy-dark matter connection.  We
showed that the analytical model presented in van den Bosch et
al. (2013) and constrained by (SDSS) abundance, clustering, and
lensing data at $z<0.2$ (see C13), reproduces, \emph{without further
  adjustments}, the galaxy-galaxy lensing signal measured around
Luminous Red Galaxies \citep{MandelCosmo12} at ${\bar z} \sim 0.26$
and ${\bar z} \sim 0.4$ and throughout the probed spatial range, $0.02
\lsim R \lsim 90 \, h^{-1}_{70}$ Mpc (see Figure~\ref{fig:modeldataLRG}).
This agreement is an important validation of the model determined by
C13. It not only implies that the parameters that describe the halo
occupation distribution are also valid at higher redshifts, but also
implies consistency with the cosmological parameters found by C13.

Following  \citet{vanUitert2011} , we measure the lensing signal
around lenses from the Sloan Digital Sky Survey (Data Release 9) using
shape measurements from the 450 square degrees that overlap with the
second generation Red-sequence Cluster Survey (RCS2, \citealt{RCS2}). The higher source density and redshift results in a
significant improvement, compared to SDSS data alone, for lenses with
$z \gsim 0.3$. We split the lenses into eight luminosity bins and
measure robust tangential shear signals as a function of the transverse
separation, $R$, in the range $0.05 \lsim R \lsim 10 h^{-1}_{70}$ Mpc
(see Figure~\ref{fig:modeldatacomparison}).

Compared to the earlier study by \citet{vanUitert2011} 
which used the overlap with DR7, the use of the overlap with DR9 increases the 
number of lenses, resulting in an improvement of about 50\% in the precision of the 
lensing shear over the entire range probed here ($0.05 \lsim R \lsim 10$ Mpc). In
addition we now include the sources from neighbouring RCS2 pointings
(previously the analysis was done on a pointing-by-pointing
basis). This increases the lensing signal-to-noise at large projected
lens-source separations (see Figure~\ref{fig:ESD-DR9}), and reduces
systematic contributions to the lensing signal (see
Figure~\ref{fig:appendix}). Finally, compared to the DR7 catalogue,
the redshift distribution of lens galaxies in DR9 has a large number
of galaxies at $z > 0.4$ (see Figure~\ref{fig:zMrhist}), enabling us
to probe the matter distribution at those high redshifts (see
Figure~\ref{fig:ESD-DR9-2}).

We split the lens galaxies in 8 luminosity bins, ranging from $-18 <
{^{0.1}M}_r < -17$ to $-23 < {^{0.1}M}_r < -22.5$. Brighter galaxies
are distributed at increasingly higher redshift such that the data
span a wide range in redshift from $\bar{z}=0.07$ to
$\bar{z}=0.59$. Moreover, since brighter galaxies live on average in
more massive haloes, the range in luminosity probed here spans a
correspondingly wide range in host halo mass. As a result, the
measurements presented here \emph{simultaneously} probe the matter
distribution in different regimes from small groups to massive
clusters, and from low to high redshift (see
Figure~\ref{fig:modeldatacomparison}).
   
Without any adjustment, the C13 model also describes the lensing signal 
obtained with RCS2 data very well.
This corroborates the results based on the SDSS analysis of LRGs, but
also implies consistency of the measurement of the lensing signal.  We
note that on the smallest scales probed here ($0.05 \lsim R \lsim 0.2
h_{70}^{-1}{\rm Mpc}$), we find better agreement if we include a
contribution from the stellar mass of the galaxies: a simple
point-mass model for the stellar component of the galaxies is
sufficient to boost model predictions at those small scales leading to
a better agreement with the data.

Finally, exploiting the high signal-to-noise ratio of the lensing
signal around bright galaxies,
we attempt to constrain aspects of the galaxy-dark matter connection
across cosmic time.
While the inference of an evolutionary scenario 
for the galaxy luminosity-halo mass relation is hampered by current 
uncertainties in the evolution of galaxy luminosity,
we robustly assess that, up to $z \sim 0.6$, the number of
central galaxies as a function of halo mass is well described by a
log-normal distribution with scatter, $\sigma_{\log L_{\rm c}}=0.146\pm0.011$, 
in agreement with previous independent studies at lower redshift.

Our results demonstrate the value of complementing the excellent
information about the properties of the lenses provided by the SDDS
with deeper, high-quality imaging data. This allows us to probe the
link between galaxies and matter around them in increasing level of
detail and at increasingly higher redshift. In this paper we tested
the model of C13 and found that it overall performs very well.
In future publications we will use our data
to examine the evolution of  early-type galaxies only, and we will carry out
a comprehensive study of the possible evolution with cosmic time of 
the galaxy luminosity-halo mass relation for early-type galaxies.

\section*{Acknowledgments}

We are grateful to Rachel Mandelbaum for providing us 
the excess surface density measurements of Luminous Red Galaxies
(Figure 1) in electronic format.  HH and MC acknowledge support from NWO VIDI grant number 639.042.814.  HH also acknowledges ERC FP7 grant 279396. 
\bibliography{paper}

\appendix

\section{The Conditional Luminosity Function}
\label{sec:clf}

Throughout the paper, the \emph{average} number of galaxies with
luminosities in the range $L \pm \rmd L/2$ that reside in a halo of
mass $M$ is described by the conditional luminosity function, $\Phi(L|M)$, 
introduced by \citet{Yang2003}:
\begin{eqnarray}
\langle N_{\rm x}|M \rangle &=&  
\int_{L_1}^{L_2} \Phi_{\rm x}(L|M) \, {\rm d} L \, .
\end{eqnarray}
Following \citet{CoorayMilo05} and \citet{Cooray2006},
we split the conditional luminosity function (hereafter CLF) in two components,
\begin{equation}
\Phi(L|M) = \Phi_\rmc(L|M) + \Phi_\rms(L|M)\,,
\end{equation}
where $\Phi_\rmc(L|M)$ describes the contribution due to central
galaxies (defined as those galaxies that reside at the center of their
host halo), while $\Phi_\rms(L|M)$ characterizes satellite galaxies
(those that orbit around a central).  

Our parameterization of the CLF model is motivated by the results
obtained by \citet{Yang2008a} from a large galaxy group
catalogue \citep{Yang2007} extracted from the SDSS Data Release 4,
and by \citet{Tal2012} from a study of the luminosity function of
satellite galaxies of luminous red galaxies.  In particular, the CLF
of central galaxies is modeled as a log-normal function:
\begin{equation}\label{phi_c}
\Phi_\rmc(L|M) \,{\rmd}L = {\log\, e \over {\sqrt{2\pi} \, \sigma_{\rm logL_c}}} 
{\rm exp}\left[- { {(\log L  -\log L_\rmc )^2 } \over 2\,\sigma_{\rm logL_c}^2} \right]\,
{\rmd L \over L}\,,
\end{equation}
and the satellite term as a modified Schechter function:
\begin{equation}\label{phi_s}
\Phi_\rms(L|M)\,{\rmd}L = \phi^*_\rms \,
\left({L\over L^*_\rms}\right)^{\alpha_\rms + 1} \,
{\rm exp} \left[- \left ({L\over L^*_\rms}\right )^2 \right] {\rmd L \over L}\,,
\end{equation}
which decreases faster than a Schechter function at the bright end.
Note that $L_\rmc$, $\sigma_\rmc$, $\phi^*_\rms$, $\alpha_\rms$ and
$L^*_\rms$ are in principle all functions of the halo mass $M$.

Following \citet{Cacciato2009}, and motivated by the results of \citet{Yang2008a} and \citet{More2009a,More2011a}, we assume that $\sigma_{\rm  logL_c}$, which expresses the scatter in $\log L$ of central
galaxies at fixed halo mass, is a constant (i.e.  is independent of
halo mass and redshift).  Note though that this does not imply that
the scatter in halo mass at a fixed luminosity, $\sigma_{\rm log M}$,
is constant: as discussed in \citet{Cacciato2009} and
\cite{More2009a}, $\sigma_{\rm log M}$ increases because the slope of
the $L_{\rm c}(M)$ relation becomes shallower with increasing $M$.  In
addition, for $L_\rmc$, we adopt the following parameterization:
\begin{equation}\label{LcM}
L_\rmc(M) = L_0 {(M/M_1)^{\gamma_1} \over 
\left[1 + (M/M_1) \right]^{\gamma_1-\gamma_2}}\,.
\end{equation}
Hence, $L_\rmc \propto M^{\gamma_1}$ for $M \ll M_1$ and $L_c \propto
M^{\gamma_2}$ for  $M \gg  M_1$. Here $M_1$  is a  characteristic mass
scale, and $L_0 =  2^{\gamma_1-\gamma_2} L_c(M_1)$ is a normalization. 
For the satellite galaxies we adopt
\begin{equation}
L^*_\rms(M)  = 0.562 L_\rmc(M)\,,
\end{equation}
\begin{equation}\label{alpha}
\alpha_\rms(M) = \alpha_\rms 
\end{equation}
(i.e., the faint-end slope of $\Phi_\rms(L|M)$ is independent of mass
and redshift), and
\begin{equation}\label{phi}
\log[\phi^*_\rms(M)] = b_0 + b_1 (\log M_{12}) + b_2 (\log M_{12})^2\,,
\end{equation}
with $M_{12}=M/(10^{12} h_{70}^{-1}\Msun)$. Note that neither of these
functional forms has a physical motivation; they merely were found to
adequately describe the results obtained by \citet{Yang2008a} from the
SDSS galaxy group catalogue.

To summarize, our parameterization of the CLF thus has a total of nine
free parameters. Based on the results of \citet{CosmoCLFIII}, 
unless otherwise specified, we adopt 
the values ($\log[M_1/(h_{70}^{-1}M_{\odot})]$, 
$\log[L_0/(h_{70}^{-2}L_{\odot})]$, 
$\gamma_1$, $\gamma_2$, 
$\sigma_{\rm logL_c}$, $\alpha_\rms$, $b_0$, $b_1$, $b_2$)=(11.39, 10.25, 3.18, 
0.245, 0.157, -1.18, -1.17, 1.53,-0.217).

\section{Shear Systematics and Covariance Matrix}
\begin{figure*}
\psfig{figure=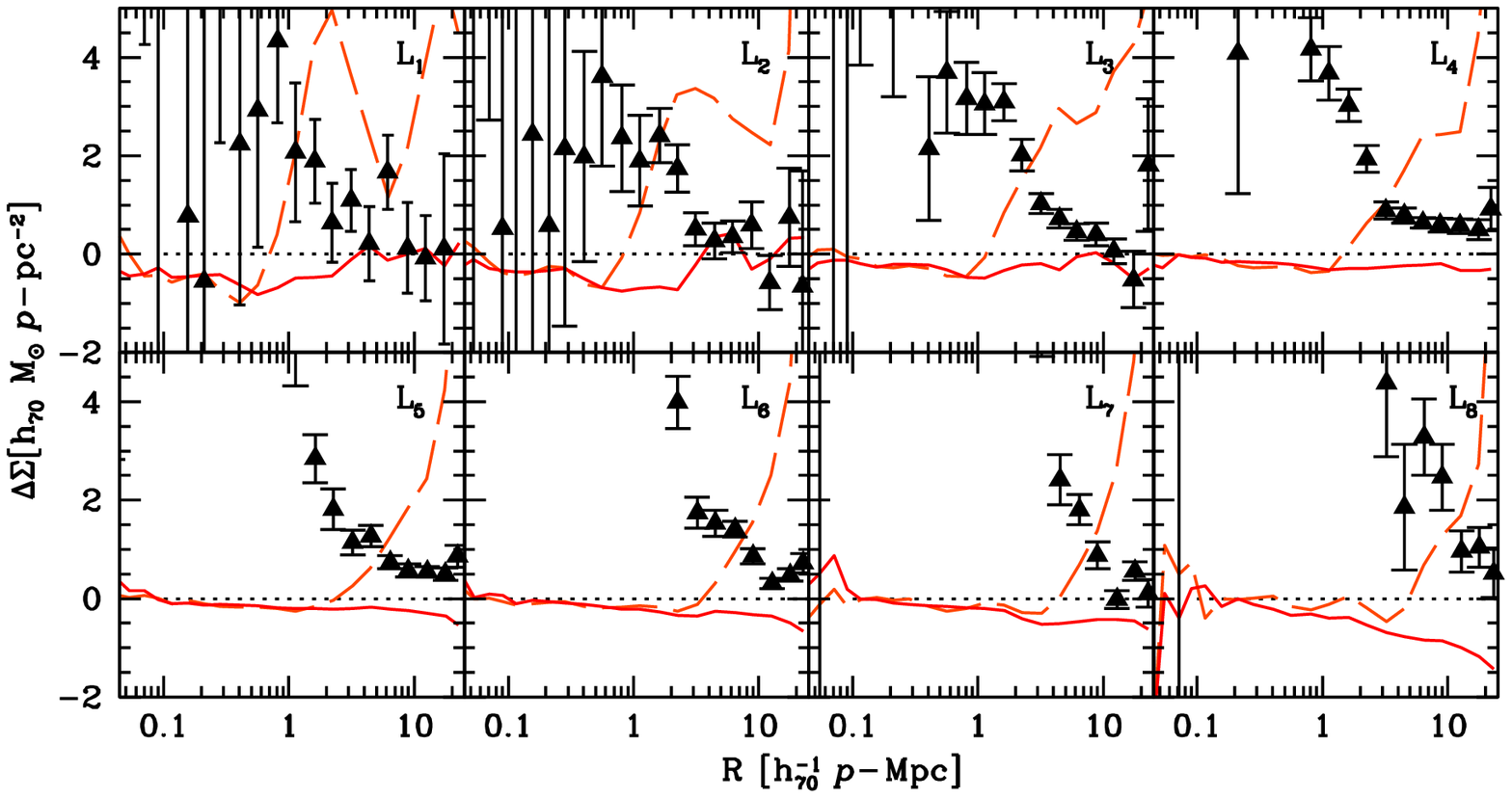,width=\hdsize}
\caption{ 
  Assessment of systematics in the excess surface density
  measurements. Each panel refer to a luminosity bin as indicated by
  the labels. Black triangles with error bars refer to the excess
  surface density signal.
  The orange dashed lines
  denote the random shear signal obtained if one were to use only one
  single exposure, whereas the red solid line refers to the random
  shear signal obtained including neighbouring patches as in our
  analysis.  }
\label{fig:appendix}
\end{figure*}

For this paper the
inaccuracies in the correction for PSF anisotropy is a main source of bias. On small scales we average over many lens-source pairs which
have a random orientation with respect to the direction of PSF
anisotropy, and as a result the lensing signal is robust. At large
radii the lensing signal is small and residual systematics may become
more dominant because the angles between the PSF anisotropy and the
lens-source pairs may no longer be isotropic because of masks or the
survey geometry. In this section we therefore examine the reliability
of the lensing signal on scales $\gsim 1 h_{70}^{-1} p-$Mpc.

To account for residual systematics in our shape measurement
catalogues, and for the impact of image masks on the lensing signal,
we compute the lensing signal around a large number of random points
and subtract that from the galaxy-galaxy lensing signal. 
In the absence of systematics or
an isotropic orientation of lens-source pairs, this signal should
vanish. 
The red line in Figure~\ref{fig:appendix} compares this signal
to the observed lensing signal (triangles with error bars) for the
various luminosity bins.  
We find that the correction is generally very small: it is 
negligible for the L1 to L5 bins, and for the other bins
it is significantly smaller than the lensing signal 
over the range $0.05<r<10$ $h_{70}^{-1} p-$Mpc which we use in our analysis. Therefore,
any small error in the calculation of the random shear signal will
have a minor effect at most on our results.

The random shear signal is small because we include neighbouring
pointings in our lensing analysis.  Consequently, at large projected
separations, the lensing signal is averaged over many more lens-source orientations, which averages out any residual systematics in
our shape measurement catalogues at those scales. That it is important
to include the neighbouring pointings, as opposed to analysing each
pointing separately, is demonstrated by the dashed orange curves: these
show the random signal for the case where we perform the lensing
measurements on single exposures only. In this case, the random signal
strongly increases with increasing lens-source separation, and its
amplitude becomes comparable or even larger than the size of the
lensing signal itself. Even a relatively small error in the
computation of the random lensing signal could seriously affect the
results, which is clearly undesirable. Therefore, it is important to
conduct the analysis on patches rather than individual exposures.

Another commonly used test is the measurement of the cross-shear,
which is the projection of the source ellipticities to a unit vector
that is rotated by 45 degrees from the vector that connects that lens
and the source. Galaxy-galaxy lensing only produces tangential shear
and not cross shear, as the average gravitational potential of a large
number of lenses with random orientations is circularly
symmetric. Therefore, if we measure a cross shear that is not
consistent with zero this indicates the presence of systematics in our
shape measurement catalogues. For all luminosity bins we find that the
cross-shear is consistent with zero on all scales used in our
analysis.

In order to quantify the level of correlation between the lensing
signals of different radial bins, we compute the correlation matrix from
the data using the ``delete one jackknife'' method (Shao 1986). We treat
each pointing of the total $N$ pointings as a sub-volume that is
subsequently left out to create a new realization of the data. The
covariance matrix is then determined as
\begin{equation}
C_{ij}(\gamma_i,\gamma_j)=\frac{N-1}{N} \sum_{k=1}^N
(\gamma_i^k-\bar{\gamma_i}) (\gamma_j^k-\bar{\gamma_j}),
\end{equation}
with $\gamma_i$ the shear of the $i$-th radial bin, $\gamma_i^k$ the
shear at that location from one of the jackknife realizations, and
$\bar{\gamma_i}$ the mean shear of that bin determined by averaging over
all the jackknife realizations. The correlation matrix follows from
${\rm Corr}_{ij}=C_{ij}/\sqrt{C_{ii}C_{jj}}$.

We find that the correlation matrix is practically diagonal for almost
all of our luminosity bins. Only for L2, L3 and L4 we find some
low-level off-diagonal terms only around scales of $\sim$1 $h_{70}^{-1} p-$Mpc.
Since our analysis is mostly sensitive to the highest luminosity bins,
we assume that the correlation matrices are diagonal when we fit the models to the data.

Note that the correlation matrix that results from the jackknife method
depends on the size of the sub-volume that is subsequently left out.
This is demonstrated in \citet{Norberg2008}, who compared several ways
to determine the variance and covariance of 2-point clustering
measurements. For our purposes, the covariance matrix we determine is
expected to be sufficiently accurate. However, for using measurements
like these to constrain cosmological parameters, this is an issue that
needs to be addressed, separately from the effect that the inverse of a
noisy but unbiased correlation matrix is not unbiased \citep{Hartlap2007}.

\section{Source Redshift Distribution}

Using lens galaxies at higher redshifts, the mean lensing
efficiency $\langle D_{\rm ls}/D_{\rm s}\rangle$ becomes more sensitive to the
adopted redshift distribution of the sources, $P(z_{\rm so})$.
Therefore, we have updated the method for determining $P(z_{\rm
so})$. \citet{vanUitert2011} used
the photometric redshift catalogues of the CFHTLS ``Deep Survey" fields
\citep{IlbertCFHTLSdeep} and selected all objects in the range $22<r'<24$
that satisfied the selection cuts as described in the release notes that
accompanied the catalogues, i.e. only objects with reliable photometry
in all the bands, that were observed in unmasked regions and with a
best-fit template number $<$ 54. Since the main interest there was to
determine the redshift distribution rather than to select galaxies with
reliable photometric redshifts, galaxies in the redshift range $0.05<z_{\rm
phot}<2.0$ were selected instead of $0.2<z_{\rm phot}<1.5$ where
the redshifts were deemed reliable. \citet{vanUitert2011} did not account for the scatter of the photometric redshifts, nor for the fraction of outliers. Also, they did not account for the fact that bright sources have a larger weight in the lensing measurements than faint ones.

To increase the precision of the lensing efficiencies for higher lens
redshifts, in this paper, we use the photometric redshift catalogue from the
2 deg$^2$ COSMOS field \citep{IlbertCOSMOS} instead. The photometry in
30 bands results in photometric redshifts that are both more accurate
than those from the CFHTLS, and also more reliable up to higher
redshifts. Using the overlap with the CFHTLS-D2 catalogue, kindly
provided by H. Hildebrandt, we determined the conversion between the
$r^+$-band from the COSMOS catalogues, and the $r'$-band from the
CFHTLS. Using this conversion we selected source galaxies in COSMOS
based on their $r^+$ magnitudes corresponding to a selection of $22<r'<24$.

When integrating $D_{\rm ls}/D_{\rm s}$ over the $P(z_{\rm so})$, we have to
account for the fact that bright galaxies have a larger weight in our
lensing measurements than faint ones. For this purpose, we determined
the average lensing weight of the source galaxies in the RCS2 in narrow
$r'$-band magnitude bins, finding that on average the sources with $r'
\sim $ 22 have a weight that is twice that of $r'\sim $ 24 source
galaxies. We used the conversion between the $r^+$- and $r'$-band to
compute the corresponding weight of each galaxy in the COSMOS catalogue,
and used that weight to determine the weighted mean lensing efficiency.

To account for the outliers, we assigned a new redshift to a random
fraction of the galaxies equal to the outlier fraction. The new redshift
was drawn from the photometric redshift distribution of the sources, and
replaced the catalogue value when it fulfilled the outlier criterion
$|z_{\rm random}-z_{\rm phot}|/(1+z_{\rm random})>0.15$. The outlier
fraction depends on the brightness; we adopted a value of 0.7\% for
galaxies with $i^+<23$, and 15.3\% for galaxies with $i^+>23$, as quoted
in \citet{IlbertCOSMOS}. We created 16 realizations of the photometric
redshift catalogues, each with a different randomly assigned set of
outliers, and adopted the mean as our new lensing efficiencies. The
scatter between the different realizations is small, and can safely be
ignored compared to the statistical errors of the lensing analysis. We
show the mean lensing efficiency at 10 lens redshifts in the second
column of Table \ref{tabpz}.

\begin{table}\label{tabpz}
  \caption{Lensing efficiencies determined from the photometric redshift
distributions of source galaxies}
\begin{center}
\begin{tabular}{lccc}
\hline\hline
 $z_{\rm lens}$ & $\langle D_{ls}/D_s\rangle$ & $\langle
D_{ls}/D_s\rangle$ (CFHT) &$\langle D_{ls}/D_s\rangle$ ($z_{\rm
phot}<2$)  \\
  (1) & (2) & (3) & (4) \\
\hline

0.1 & 0.774 & 0.777 (1.00)& 0.768 \\
0.2 & 0.602 & 0.586 (1.03)& 0.588 \\
0.3 & 0.461 & 0.440 (1.05)& 0.443 \\
0.4 & 0.350 & 0.328 (1.07)& 0.329 \\
0.5 & 0.264 & 0.240 (1.10)& 0.241 \\
0.6 & 0.194 & 0.172 (1.13)& 0.172 \\
0.7 & 0.141 & 0.118 (1.19)& 0.118 \\
0.8 & 0.103 & 0.079 (1.30)& 0.081 \\
0.9 & 0.076 & 0.053 (1.43)& 0.055 \\
1.0 & 0.057 & 0.036 (1.58)& 0.037 \\

\hline\hline
\end{tabular}
\end{center}
\medskip
\begin{minipage}{\hssize}
  (1) lens redshift; (2) lensing efficiency determined using the
photometric redshift catalogues of COSMOS \citep{IlbertCOSMOS}; (3)
lensing efficiency determined using the photometric redshift catalogues
of the CFHTLS ``Deep Survey" fields \citep{IlbertCFHTLSdeep}, restricted to
source galaxies in the range $z_{\rm phot}<2$. The bracketed values show
the ratio between column 2 and 3; (4) lensing efficiency determined
using the photometric redshift catalogues of COSMOS, restricting the source galaxies in the range $z_{\rm phot}<2$.
\end{minipage}
\end{table}

We ignored the impact of scatter of the photometric redshifts with
respect to the spectroscopic redshifts. The effect of scatter is that it
moves galaxies in redshift from where their abundance is large to where
it is small. To estimate the impact that might have on $\langle
D_{\rm ls}/D_{\rm s}\rangle$, we additionally scattered each photometric redshift
by randomly drawing a value from a Gaussian, whose width depends on the
galaxies' $i$-band magnitude, as quoted in \citet{IlbertCOSMOS}. We
multiplied that random value with $1+z_{\rm phot}$ and added it to
$z_{\rm phot}$\footnote{Formally, \citet{IlbertCOSMOS} quote the
scatter on $\Delta z/(1+z_{\rm spec})$ with $\Delta z = z_{\rm
phot}-z_{\rm spec}$, so we should have multiplied the random value with
$1+z_{\rm spec}$ rather than $1+z_{\rm phot}$. However, we expect the
difference to be minor.}. We created 16 new realizations, and determined
the mean lensing efficiency. We found that the impact is less than a
percent at all lens redshifts, and can therefore be safely ignored.

To see how the lensing efficiencies compare to those computed using the
CFHTLS ``Deep'' catalogues, we applied the same procedure to compute the
average lensing efficiencies. We accounted for outliers by adopting the
outlier fractions as a function of $i'$-band magnitude from Ilbert et
al. (2006), and applied the same weight as a function of $r'$-band
magnitude. However, we only selected galaxies with $z_{\rm phot}<2$.
Again, we created 16 realizations, and determined the mean. We show the
resulting values of $\langle D_{\rm ls}/D_{\rm s}\rangle$ in the third column of
Table \ref{tabpz}. At low redshifts, the resulting lensing efficiencies
only differ by a few percent compared to the ones based on the COSMOS
catalogue. However, we find that if the lens redshift increases, the
$\langle D_{\rm ls}/D_{\rm s}\rangle$ from COSMOS becomes increasingly larger. 
To demonstrate that this difference is due to source galaxies at $z_{\rm
phot}>2$, we repeated the calculation using the COSMOS photometric
redshift catalogue, but now restricting the analysis to $z_{\rm
phot}<2$. We show the resulting lensing efficiencies in the fourth
column of Table \ref{tabpz}. We find that the lensing efficiencies agree
very well with those based on the CFHTLS ``Deep'' catalogues. The
difference is at most 4\% over the entire redshift range that we probed.

In previous work where we used the photometric redshift catalogues from
\citet{IlbertCFHTLSdeep} to compute the lensing efficiencies, we focused at
galaxies at low redshifts. Hence the lensing efficiencies that we used
there were of sufficient accuracy. However, for galaxies at redshifts
$z>0.5$, our results show that is it important to include source
galaxies at $z_{\rm phot}>2$ in the computation of $\langle
D_{ls}/D_s\rangle$.

Note that we have ignored cosmic variance. However, we find very similar
lensing efficiencies using the COSMOS and CFHTLS ``Deep'' photometric redshift catalogues when we restrict the galaxies to $z_{\rm phot}<2$. This
suggests that cosmic variance does not have a large impact on the
lensing efficiencies that we use.
\label{lastpage}
\end{document}

%% file: macros.tex
\def\beq{\begin{equation}}
\def\eeq{\end{equation}}
\def\barray{\begin{eqnarray}}
\def\earray{\end{eqnarray}}
\def\beqarray{\begin{eqnarray}}
\def\eeqarray{\end{eqnarray}}

\def\rmb{{\rm b}}
\def\rmc{{\rm c}}
\def\rmd{{\rm d}}

\def\rmg{{\rm g}}
\def\rmh{{\rm h}}

\def\rmm{{\rm m}}

\def\rms{{\rm s}}

\def\rmx{{\rm x}}
\def\rmy{{\rm y}}

\def\calH{{\cal H}}

\newcommand{\kmsmpc}{\>{\rm km}\,{\rm s}^{-1}\,{\rm Mpc}^{-1}}

\newcommand{\Msun}{\>{\rm M_{\odot}}}

\def\gtsima{$\; \buildrel > \over \sim \;$}
\def\ltsima{$\; \buildrel < \over \sim \;$}
\def\prosima{$\; \buildrel \propto \over \sim \;$}
\def\gsim{\lower.7ex\hbox{\gtsima}}
\def\lsim{\lower.7ex\hbox{\ltsima}}
\def\simgt{\lower.7ex\hbox{\gtsima}}
\def\simlt{\lower.7ex\hbox{\ltsima}}
\def\simpr{\lower.7ex\hbox{\prosima}}

\newdimen\hssize
\newdimen\hdsize
